# Electronic structure of $Cr_{1-\delta}S$ ($\delta$=0, 0.17) with NiAs-type crystal structure

M. Koyama[a], H. Sato[b,*], Y. Ueda[a], C. Hirai[b] and M. Taniguch[b,c]

[a]*Kure National College of Technology, Aganinami 2-2-11, Kure 737-8506, Japan*
[b]*Gradiate School of Science, Hiroshima University, Kagamiyama 1-3-1, Higashi-Hiroshima 739-8526, Japan*
[c]*Hiroshima Synchrotron Radiation Center, Hiroshima University, Kagamiyama 2-313, Higashi-Hiroshima 739-8526, Japan*

**ABSTRACT**

Valence-band and conduction-band electronic structure of CrS ($\delta$=0) and $Cr_5S_6$ ($\delta$=0.17) has been investigated by means of photoemission and inverse-photoemission spectroscopies. Bandwidth of the valence bands of $Cr_5S_6$ (8.5 eV) is wider than that of CrS (8.1 eV), though the Cr 3d partial density of states evaluated from the Cr 3p-3d resonant photoemission spectroscopy is almost unchanged between the two compounds with respect to the shapes including binding energies. The Cr 3d ($t_{2g}$) exchange splitting energies of CrS and $Cr_5S_6$ are determined to be 3.9 and 3.3 eV, respectively.

It is well known that chromium chalcogenides based on the NiAs-type structure exhibit various physical properties depending on the chalcogen and ordered Cr vacancy [1-6]. $Cr_{1-\delta}Te$ are ferromagnetic metals with the $\delta$-dependent Currie temperatures, while $Cr_{1-\delta}Se$ and $Cr_{1-\delta}S$ exhibit predominantly antiferromagnetic behavior [7]. Among the chromium chalcogenides, CrS is an antiferromagnet with the Néel temperature of 450 K, while $Cr_5S_6$ a ferrimagnet with the Curie temperature of 300 K and shows the antiferromagnetic behavior below 150 K [2]. The metal-semiconductor transition is observed for $Cr_{1-\delta}S$ with $\delta\leq0.1$ at ~620 K, while $Cr_{1-\delta}S$ with $\delta>0.1$ are metals. The crystal structures of CrS and $Cr_5S_6$ are shown in Fig. 1, where the Cr vacancies of $Cr_5S_6$ order regularly and the unit cell becomes twice of that of CrS along the c-axis.

A wide variety of physical properties of $Cr_{1-\delta}S$, which change also with the applied pressure, is derived from the electronic structure with respect to the Cr 3d states and hybridization between the Cr 3d and S 3p states. Some band-structure calculations have been carried out on CrS so far [8-10]. However, there is no experimental information on electronic structure of $Cr_{1-\delta}S$.

In this study, we report the valance-band and conduction-band densities of states (DOSs) of $Cr_{1-\delta}S$ ($\delta$=0, 0.17) obtained by means of the ultraviolet photoemission and inverse-photoemission spectroscopies (UPS and IPES) as well as the Cr 3d partial DOSs of $Cr_{1-\delta}S$ deduced from the resonant photoemission spectroscopy (RPES) in the Cr 3p-3d excitation region. These DOSs are compared with the result of the band-structure calculation and discussed in the light of the effect of the vacancies at the Cr site on electronic structure.

Samples used for the present experiments were polycrystals of $Cr_{1-\delta}S$ ($\delta$=0, 0.17) prepared by the following procedure. An appropriate amount of Cr (99.999%) and S (99.999%) elements sealed in an evacuated quartz ampoule was heated at 1173 K for a week. The reacted products were crushed and mixed well in Ar atmosphere. The products sealed in the evacuated ampoule were again heated at 1773 K for several hours, cooled down to 973 K in 24 hours and then quenched into iced water. The crystal structure and composition were checked by the x-ray powder diffraction and electron-probe micro-analysis (EPMA), respectively [11]. Lattice parameters of CrS and $Cr_5S_6$ with single phase of the NiAs-type structure at room temperature are $a_0$=3.469 Å and $c_0$=5.786 Å, and $a_0$=3.448 Å and $c_0$=5.751 Å, respectively, where $a_0$ and $c_0$ are defined in Fig. 1. The smaller lattice parameters of $Cr_5S_6$ are noticed due to the Cr vacancies. Clean surfaces for the UPS, IPES and RPES measurements were *in situ* obtained by scraping with a diamond file.

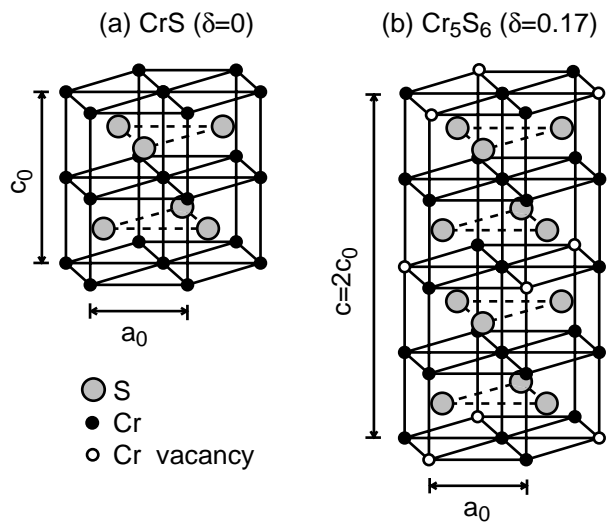

**Figure 1.** Crystal structures of (a) CrS and (b) $Cr_5S_6$.

Apparatus for UPS and IPES is composed of three chambers for sample preparation, UPS experiments and IPES experiments with base pressures of $2\times10^{-10}$, $2\times10^{-10}$ and $5\times10^{-10}$ Torr, respectively. The UPS spectrometer is made up of an He discharge lump ($h\nu$=21.2 eV) and a double-cylindrical mirror-analyzer (DCMA). Pass energy of photoelectrons through DCMA was set to 16.0 eV with the corresponding energy resolution of 0.20 eV. The IPES spectrometer consists of an Erdmann-Zipf low-energy electron gun and a bandpass photon detector centered at $h\nu$=9.43 eV [12,13]. Total energy resolution is 0.56 eV. The energy of the UPS and IPES spectra was defined with respect to the Fermi level ($E_F$) and both spectra were connected at $E_F$.

The Cr 3p-3d RPES experiments were carried out on the beamline BL7 at Hiroshima Synchrotron Radiation Center (HSRC). Light from the storage ring (HiSOR) is monochromatized by the Dragon-type monochromator [14]. We used a photoemission spectrometer with a hemispherical photoelectron analyzer (Gammadata Scienta SES100) attached to the end station of BL7. The total energy resolution was set below 0.1 eV around $h\nu$=50 eV. All experiments were performed at room temperature.

Figure 2 (a) shows the valence-band UPS spectra measured at $h\nu$=21.2 eV and the conduction-band IPES spectra of CrS ($\delta$=0) and $Cr_5S_6$ ($\delta$=0.17). The UPS spectrum of CrS exhibits a weak shoulder at -0.7 eV, two peaks at -1.5 and -6.7 eV and a broad structure around -4.7 eV between the two peaks. The IPES spectrum, on the other hand, shows a structure at 2.4 eV and a broad structure between 6 and 11 eV. The whole features of the UPS and IPES spectra of $Cr_5S_6$ are similar to those of CrS. We find a weak shoulder at -0.7 eV, peak structures at -1.5, -4.7, -6.4 and 1.8 eV and a broad structure between 6 and 11 eV in the spectra of $Cr_5S_6$.

Figure 2 (b) shows theoretical total DOS for the antiferromagnetic phase of CrS by means of the band-structure calculation using the augmented spherical wave method [8]. The peak structures appear at -7.1, -4.7, -0.7 and 1.9 eV. The hybridization between the Cr 3d and S 3p states exists over the calculated energy region. In the hybridization bands, the Cr 3d ↑ and Cr 3d ↓ states with $t_{2g}$ symmetry mainly contribute to the peaks at -0.7 and 1.9 eV, respectively, providing the Cr 3d ($t_{2g}$) exchange splitting energy of 2.6 eV. The S 3p states hybridized with the Cr 3d states with $e_g$ symmetry, on the other hand, contribute to the peak at -4.7 eV. The peak at -7.1 eV is predominantly derived from the S 3p states.

For the valence bands of $Cr_{1-\delta}S$ mainly composed of the Cr 3d and S 3p states, the photoionization cross-section of the Cr 3d state is only twice lager than that of the S 3p state at $h\nu$=21.2 eV [15]. The UPS spectra of $Cr_{1-\delta}S$ measured at $h\nu$=21.2 eV, thus, provide information on the Cr 3d states as well as the S 3p states. The experimental results suggest that the Cr 3d - S 3p hybridization bands spread over the top ~8 eV region in the valence bands. The whole features of the UPS and IPES spectra are in good agreement with the theoretical DOS. On the basis of the band-structure calculation, we assign the two peaks near $E_F$ in the UPS and IPES spectra of $Cr_{1-\delta}S$ to the occupied Cr 3d ($t_{2g}$) ↑ and unoccupied Cr 3d ($t_{2g}$) ↓ states with nearly localized character, respectively. The structures around -4.7 eV in the UPS spectra are attributed to the S 3p states hybridized with the Cr 3d ($e_g$) ↑ states, while the most deeper peaks at -6.7 eV for CrS and at -6.4 eV for $Cr_5S_6$ are almost derived from the S 3p states.

Although the UPS and IPES spectra of $Cr_{1-\delta}S$ are similar, we find several differences with respect to the following points; (i) The bandwidth of the valence bands of $Cr_5S_6$ (8.5 eV) are wider than that of CrS (8.1 eV). (ii) The energy position of the S 3p-derived peak of CrS shifts toward the $E_F$ side by 0.3 eV for $Cr_5S_6$. (iii) The Cr 3d ($t_{2g}$) exchange splitting energy of $Cr_5S_6$ (3.9 eV) is larger than that of CrS (3.3 eV), which is larger than the theoretical value of 2.6 eV. It is noted that the energy position of the Cr 3d ($t_{2g}$) ↑ peak is at -1.5 eV for both compounds in spite of the energy shift of the S 3p-derived peak.

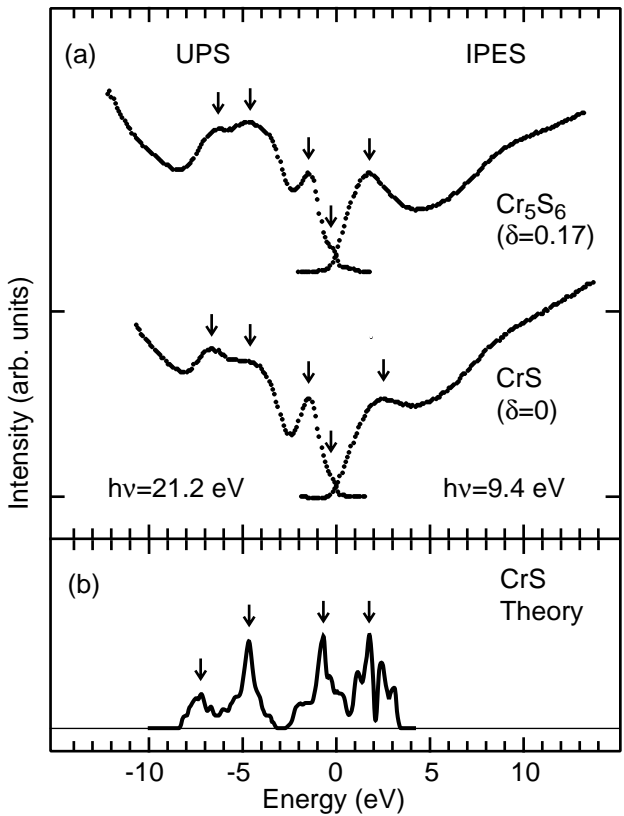

**Figure 2.** (a) UPS spectra measured at $h\nu$=21.2 eV and IPES spectra of $Cr_{1-\delta}S$. (b) The total DOS of CrS derived by the band-structure calculation [8].

In order to deduce the Cr 3d contributions to the valence bands of $Cr_{1-\delta}S$, we have carried out the RPES experiments in the Cr 3p-3d excitation region. Figures 3 (a) and (b) exhibit a series of the RPES spectra of CrS for $h\nu$ between 38.3 and 60.0 eV and the Cr 3p-3d core absorption spectrum measured by means of total electron yield, respectively. Photoemission intensities have been normalized to the monochromator output. Vertical bars indicate structures due to emissions of the Cr MVV Auger electrons. One can recognize that the peak at -1.5 eV exhibits the prominent resonance. With increasing $h\nu$ from 38.3 eV, the intensity of the main peak at -1.5 eV decreases to the minimum at $h\nu$=40.8 eV, reaches sharply to the maximum at $h\nu$=47.5 eV, and then decreases gradually with $h\nu$. The resonance enhancement takes place as a result of an interference between the direct excitation process of the Cr 3d electrons ($3p^63d^4 + h\nu \rightarrow 3p^63d^3 + \varepsilon_f$) and the discrete Cr 3p-3d core excitation process followed by a super Coster-Kronig decay ($3p^63d^4 + h\nu \rightarrow 3p^53d^5 \rightarrow 3p^63d^3 + \varepsilon_f$), where $\varepsilon_f$ represents emitted photoelectrons. Since only the Cr 3d states are resonantly enhanced for $h\nu$ near the Cr 3p-3d excitation region, we can estimate a measure of the Cr 3d contribution to the valence bands (Cr 3d partial DOS) by subtracting the spectrum at antiresonance ($h\nu$=40.8 eV) from that on resonance ($h\nu$=47.5 eV).

The derived Cr 3d partial DOSs of $Cr_{1-\delta}S$ are shown in Fig. 4, where the background contributions due to the secondary electrons have been removed. The Cr 3d DOS is composed of a weak shoulder near $E_F$, a main peak at -1.5 eV and a broad structure between -2.5 and -6 eV. The features of the Cr 3d DOSs of $Cr_{1-\delta}S$ are the same with respect to the peak-energy position and relative intensity. The RPES results again support the assignment of the peak at -1.5 eV in the UPS spectra to the Cr 3d ($t_{2g}$) ↑ states. It should be noticed that almost no contribution of the Cr 3d states to the valence-band region deeper than -6 eV. The Cr 3d - S 3p hybridization bands are in the top 6 eV, and the peaks at -6.7 eV (CrS) and -6.4 eV ($Cr_5S_6$) are derived from the S 3p states.

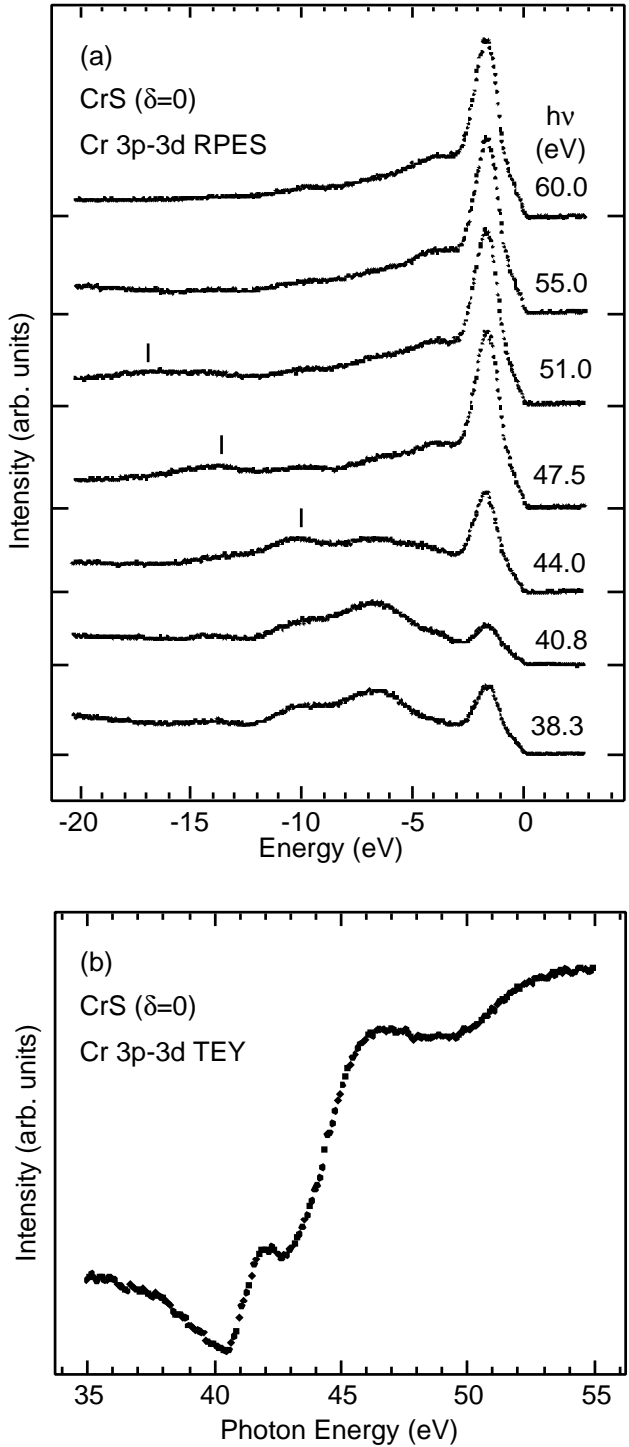

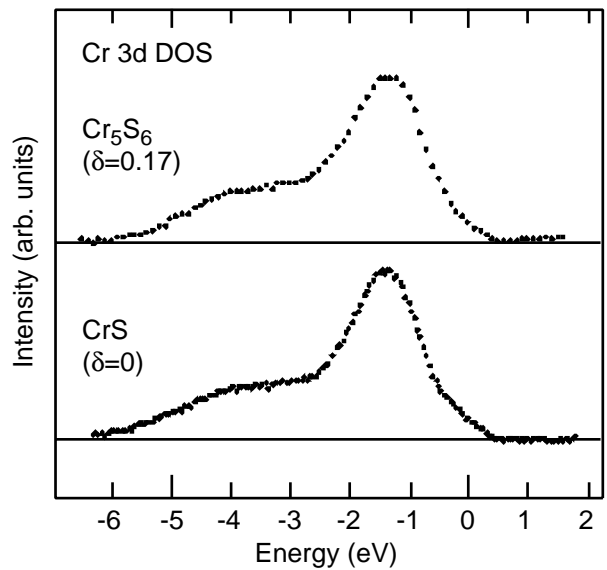

**Figure 3.** (a) A series of photoemission spectra of CrS in the Cr 3p-3d excitation region. The photoemission intensities have been normalized to monochromator output. (b) The Cr 3p-3d core absorption spectrum measured by means of tota electron yield.

**Figure 4.** Cr 3d partial DOSs of CrS and $Cr_5S_6$. Background contribution due to the secondary electrons has been removed.

Here, we discuss the Cr vacancy-dependence of the electronic structure of $Cr_{1-\delta}S$. The Cr ion has nominally $3d^4$ electron configuration in the CrS ($Cr^{2+}S^{2-}$) compound. With an increase of $\delta$, the number of electrons occupying the Cr 3d orbitals decreases and becomes $3d^3$ configuration in the $Cr_2S_3$ compounds under the assumption that the valence of the S ions remains 2-. Hole doping into the Cr 3d bands due to the Cr vacancies would lead to the energy shift of the Cr 3d-derived peak in the photoemission spectra toward the $E_F$ side with $\delta$, in contradiction to the experimental results in Figs. 2 (a) and Fig. 4, where the energy position of the Cr 3d ($t_{2g}$)-derived peak is unchanged at -1.5 eV between the two compounds. The energy shift by ~0.3 eV toward the $E_F$ side is observed for the S 3p-derived peaks from CrS ($\delta$=0, -6.7 eV) to $Cr_5S_6$ ($\delta$=0.17, -6.4 eV), rather than for the Cr 3d peaks. This suggests that the holes are mainly introduced to the S 3p bands in the $Cr_{1-\delta}S$ compounds.

Recently, we have investigated magnetic circular dichroism (MCD) in the Cr 2p-3d absorption spectra of $Cr_{1-\delta}Te$ [16]. Theoretical analyses for the MCD spectra using a configuration interaction theory with a $CrTe_6$ cluster also suggest that holes are mainly doped into the Te 5p states rather than the Cr 3d states.

The wider bandwidth of the Cr 3d - S 3p hybridization bands and the smaller Cr 3d ($t_{2g}$) exchange splitting energy of $Cr_5S_6$ than those of CrS are qualitatively explained by taking into account the smaller lattice constants of $Cr_5S_6$ caused by existence of the Cr vacancies. The difference of the bandwidth is understood as originating from the higher degree of the Cr 3d - S 3p hybridization in $Cr_5S_6$ in comparison with that in CrS. As concerns the Cr 3d ($t_{2g}$) exchange splitting energy, the Cr 3d states directly interact with the nearest Cr 3d states, since the Cr ions order in the c plane in the NiAs-type structure (see Fig. 1), and the Cr 3d - Cr 3d hybridization plays an important role in determining the Cr 3d ($t_{2g}$) exchange splitting energy [17]. The smaller Cr 3d ($t_{2g}$) exchange splitting energy for $Cr_5S_6$ than that for CrS is assumed to be due to the higher degree of the Cr 3d - Cr 3d hybridization in $Cr_5S_6$ in comparison with that in CrS.

In summary, the UPS and IPES spectra of $Cr_{1-\delta}S$ suggest that the hole due to the ordered Cr vacancies are mainly doped into the S 3p bands rather than the Cr 3d states. The wider bandwidth of the valence bands and the smaller Cr 3d ($t_{2g}$) exchange splitting energy of $Cr_5S_6$ are qualitatively explained on the basis of the smaller lattice constants of $Cr_5S_6$.

**Acknowledgements**

The authors are grateful to K. Takada and H. Okuda for their experimental supports.

**References**

[1]  F. Jellinek, *Acta Cryst.* **10**, 620(1957).

[2]  T. Kamigaichi, *J. Sci. Hiroshima Univ.* **A24**, 371 (1960) .

[3]  K. Masumoto, T. Kamigaichi, T. Hihara, *J. Phys. Soc. Jpn.* **15**, 1355 (1960).

[4]  T.J.A. Popma, C.F.Van Bruggen, *J, Inorg. Nucl. Chem.* **31**, 73 (1969).

[5]  H. Konno, M.Yuzuri, *J. Phys. Soc. Jpn.* **57**, 621 (1988).

[6]  V. V. Sokolovich, *Sov. Phys. Solid State* **34**, 371 (1992).

[7]  J. B. Goodenough, *Magnetism and the Chemical Bond*, Interscience, New York, 1963.

[8]  J. Dijkstra, C. F. van Bruggen, C. Haas, R. A. de Groot, *J.Phys. : Condens. Matter* **1**, 9163 (1989).

[9]  T. Kawakami, N. Tanaka, K. Motizuki, *J. Magn. Magn. Mater.* **196-197**, 629 (1999).

[10] D. Hobbs, J. Hafner, *J. Phys. : Condens. Matter* **11**, 8197 (1999).

[11] Although the mixed NiAs-type and monoclinic phases for CrS has been reported [6], the reflections from the monoclinic phase were not observed in the present x-ray diffraction patterns.

[12] K. Yokoyama, K. Nishihara, K. Mimura, Y. Hari, M. Taniguchi, Y. Ueda, M. Fujisawa, *Rev. Sci. Instrum.* **64**, 87 (1993).

[13] Y. Ueda, K. Nishihara, K. Mimura, Y. Hari, M. Taniguchi, M. Fujisawa, *Nucl. Instrum. Methods* **A330**, 140 (1993).

[14] M. Taniguchi, J. Ghijsen, *J. Syn. Rad.* **5**, 1176 (1998).

[15] J. J. Yeh, I. Lindau, *At. Data Nucl. Data Tables* **32**, 1 (1985).

[16] K. Yaji *et al.*, unpublished.

[17] H. Sato, M. Taniguchi, K. Mimura, S. Senba, H. Namatame, Y. Ueda, *Phys. Rev.* **B61**, 10622 (2000).